# Superstatistics and Lifetime

**V. V. Ryazanov**

E-mail: vryazan@kinr.kiev.ua

Institute for Nuclear Research Ukrainian Academy of Science, 47, Nauki Avenue, Kiev, 03068, Ukraine

**Absrtact**

To describe the nonequilibrium states of a system we introduce a new thermodynamic parameter - the lifetime of a system. The statistical distributions which can be obtained out of the mesoscopic description characterizing the behaviour of a system by specifying the stochastic processes are written. Superstatistics, introduced in [1] as fluctuating quantities of intensive thermodynamical parameters, are obtained from statistical distribution with lifetime (random time to system degeneracy) as thermodynamical parameter (and also generalization of superstatistics). Necessary for this realization condition with expression for average lifetime of equilibrium statistical system obtained from stochastical storage model [25] is consist. The obtained distribution passes in Gibbs or in superstatistics distribution depending on a measure of dissipativity in the system.

## 1. Introduction

In [1] the generalization of Boltzmann factor $exp\{-\beta_0 E\}$ was introduced in the following form:

$$B(E) = \int_0^\infty d\beta` f(\beta`) exp\{-\beta` E\}. \qquad (1)$$

It is supposed there that the intensive parameter (return temperature, chemical potential, etc.) fluctuates. These fluctuations evolve on a long time scale. Locally, in some spatial area (cell) where $\beta$ is approximately constant, the system is described by usual Boltzmann-Gibbs statistics with ordinary Boltzmann factor $exp\{-\beta E\}$ where $E$ is the energy a microstate associated with each cell. On a long time scale it is necessary to take into account fluctuations of $\beta$. Superposition of two statistics (that of $\beta$ and that of $exp\{-\beta E\}$) which therefore and refers to as "superstatistics" is derived. This formalism is successfully applied to the description of fully developed hydrodynamic turbulence, defect motion in convection states, the statistics of cosmic rays and other metastable and nonequilibrium phenomena. The special case of these superstatistics, at function $f$, reduces to gamma-distribution, appears in the nonextensive statistical mechanics [2, 3], describing a number of the physical phenomena which are not satisfactory described by Boltzmann-Gibbs statistics (for example, long-range many body systems, systems with memory, many phenomena in nuclear physics, astrophysics, geophysics, ecology, the social – communities and other complex systems). In the present work the superstatistics as (1) (together with its generalization) is obtained starting from nonequilibrium thermodynamics which as a thermodynamic variable contains a lifetime of statistical system [4, 5], section 2.

## 2. System lifetime and lifetime distribution

Characterizing the nonequilibrium state by means of an additional parameter related to the deviation of a system from the equilibrium (field of gravity, electric field for dielectrics etc) was used in [6]. In the present paper we suggest a new choice of such an additional parameter as the lifetime of a physical system which is defined as a first-passage time till the random process $y(t)$ describing the behaviour of the macroscopic parameter of a system (energy, for example) reaches its zero value. The lifetime is thus a random process which is slave (in terms of the definitions of the theory of random processes [7]) with respect to the master process $y(t)$,

$$\Gamma_x = \inf\{t: y(t)=0\}, \quad y(0)=x>0 . \qquad (2)$$

The characteristics of $\Gamma$ depend on those of $y(t)$. Introducing $\Gamma$ means effective account for more information than merely in linear terms of the canonical distribution. It is important to make clear a physical interpretation to the definition (2). So, the lifetime is related to the period of stable existence of a system, its time of dwelling within the homeokinetic plateau whose distribution was related in [8] to the entropic and information parameters of a system, its response to internal and external influences, its stability and adaptation facilities. The states of a system within the homeokinetic plateau are characterized by the mutual compensation of the entropic effects related to the energy dissipation and by the effects of negative entropy determined by the existence of the negative feedbacks. When the system exits out of the limits of the plateau unstable structures and sharp qualitative changes in the behaviour of the system arise. We assume that the existence and the magnitude of finite lifetime is related to the deviation of the system from the equilibrium. The solution to the lifetime problem stands close to the problem of Kramers [9] (overcoming the potential barrier).

In the works on statistical physics similar definitions of thermodynamic values are encountered. For example, in [10]: "Any function $B(z)$ of dynamic variables ($z=q_1,...,q_N, p_1,...,p_N$), having macroscopical character, by definition is random internal thermodynamic parameter". In [6] all values dependent on $z$ concern are treated in this way. The fact that a lifetime (2) $\Gamma(z)$ is the function of $z$, is obvious from the equations for distribution of a lifetime in Markov model , which are related to the equation the $z$-dependent density of distribution. Evidently it is visible from a situation when one considers as system some volume with gas in which all molecules are located on its borders with the velocities directed outwards. Then a lifetime is finite. For other configuration of coordinates and pulses a lifetime will be another. To validate the choice of lifetime as thermodynamic value one can ressort to the method of nonequilibrium statistical Zubarev operator (*NSO*) [11] which is interpreted in [12] as averaging of quasiequilibrium (relevant) the statistical operator on distribution of lifetime of the system. Then at $p_q(y) = \varepsilon \exp\{-\varepsilon y\}$

$$ln\mathbf{r}(t)= \int_0^\infty p_q(y)ln\mathbf{r}_q(t-y,-y)dy=ln\mathbf{r}_q(t,0)- \int_0^\infty exp\{-\mathbf{e}y\}(dln\mathbf{r}_q(t-y,-y)/dy)dy \ . \qquad (3)$$

The entropy production operator [11] is equal $\mathbf{s}(t-u,-u)=dln\mathbf{r}_q(t-u,-u)/du$. If $\mathbf{s}(t-u,-u)\gg\mathbf{s}(t)$ or $\mathbf{s}(t-u,-u)\gg\mathbf{s}(t)$ has weak $u$-dependence then equality (3) acquires a form $ln\mathbf{r}(t)=ln\mathbf{r}_q(t,0)+<\mathbf{s}(t)>\mathbf{e}^{-1}=ln\mathbf{r}(t)=ln\mathbf{r}_q(t,0)+<\mathbf{s}(t)><\mathbf{G}>$, as in interpretation [12] $\mathbf{e}^{-1}=<\mathbf{G}>=<t-t_0>$ is average lifetime. We shall note, that such distribution is received in [13]. Then a relation (3) in nonequilibrium thermodynamics with lifetime as thermodynamical parameter is replaced with expressions (5) - (6), or, in more general form, $ln\mathbf{r}(t)=ln\mathbf{r}_q(t,0)+ln[Z(\mathbf{b})/Z(\mathbf{b},\mathbf{g})]-\mathbf{g}\mathbf{G}$, where $\mathbf{r}=exp\{-\mathbf{b}E-\mathbf{g}\mathbf{G}\}/Z(\mathbf{b},\mathbf{g})$; $\mathbf{r}_q=exp\{-\mathbf{b}E\}/Z(\mathbf{b})$ (in (5) instead of $<\mathbf{G}>(dln\mathbf{r}_q(t-y_1,-y_1)/dt)$ stand the random value $\mathbf{G}$ and Lagrange multiplier $\mathbf{g}$ in: $\mathbf{g}\mathbf{G}+lnZ(\mathbf{b},\mathbf{g})/Z(\mathbf{b}))$. That is, averaging on distribution of lifetime in Zubarev *NSO* is replaced with use of a random variable of lifetime, - there lays essential difference of distribution (5) - (6) from (3) and from results of works [11,13].

The value $\mathbf{G}_x$ in (2) it is time before degeneration, destructions of the system. From the theory of the random processes it follows that existence and finiteness of the lifetime $\mathbf{G}$ value is related to the presence of stationary states, which physically corresponds to the existence of stationary structures. Complex functional and hierarchical relations of real systems correspond to the analogous between livetimes determining evolutionary processes as sequence of transitions between various classes of the system states. In contrast to the traditional representation about time as about something changeable, lifetime acts as result of existence of stable stationary structures. It depends on external influences on system and from internal interactions in it.

In termostatics it is supposed, that any isolated thermodynamic system eventually reaches an equilibrium state and never spontaneously leaves it. However the assumption of a possibility of fluctuations in system (i.e. random deviations of internal parameters from their equilibrium values) contradicts main principles termostatics. We shall replace the assumption of a system remaining in the basic (equilibrium) state for an infinite long time at its isolation from influence of environment with a more physical assumption (corresponding also to the statistical description) about an opportunity of a system leaving this state and destruction, degeneration of the system under influence of internal fluctuations. The open system is considered, dynamical values under influence of interaction with an environment become random; probably only macroscopical description.

The distribution for lifetime $\mathbf{G}$ (2) generally depends on macroscopical value $y(t)$ of the master process whith respect to $\mathbf{G}$. Let us suppose that the process $y(t)=E$ is the energy of a system (equivalently one could choose the particle number, pulse etc). Gibbs equilibrium distribution for microcanonic probability density in phase space $z$ corresponds to a condition of equiprobability of all possible microstates compatible with the given value of a macrovariable. We shall assume, that transition of system in a nonequilibrium condition breaks equality of probabilities, characteristical for an equilibrium case. One introduces additional observable macroparameter, thus extending the phase space (containing additional degenerate, absorbing states). We shall assume validity of a principle of equal probabilities for the extended phase space divided into cells with constant values $(E, \mathbf{G})$ (instead of phase cells with constant values $E$). The standard procedure (e.g. [14]) allows one to write down the relation between the distribution density $P(E,\mathbf{G})=p_E\mathbf{d}(x,y)$ and microscopic (coarse-grained) density $\mathbf{r}(z;E,\mathbf{G})$

$$P(E,\mathbf{G})= \int\mathbf{d}(E-E(z))\mathbf{d}(\mathbf{G}-\mathbf{G}(z))\mathbf{r}(z;E,\mathbf{G})dz= \mathbf{r}(z;E,\mathbf{G})\mathbf{w}(E,\mathbf{G}). \qquad (4)$$

The structure factor $w(E)$ is thus replaced by $w(E,G)$ - the volume of the hyperspace containing given values of $E$ and $G$. If $n(E,G)$ is the number of states in the phase space which have the values of $E$ and $G$ less than given numbers, then $w(E,G)=d^2n(E,G)/dEdG$. It is evident that $\int w(E,G=y)dy=w(E)$. The number of phase points between $E,E+dE$; $G,G+dG$ equals $w(E,G)dEdG$. We make use now of the principle of equiprobability applied to the extended cells $(E,G)$.

Using a maximum-entropy principle [15], it is possible to write down expression for microscopic (but coarse-grained) density of probability in the extended phase space

$$r(z;E,G)=exp\{-bE-gG\}/Z(b,g), \qquad (5)$$

where

$$Z(b,g)=\int exp\{-bE-gG\}dz=\int\int dEdG\,w(E,G)exp\{-bE-gG\} \qquad (6)$$

is the partition function, $b$ and $g$ are Lagrange multipliers satisfying the equations for the averages

$$<E>=-\partial lnZ/\partial b|_g; \qquad <G>=-\partial lnZ/\partial g|_b. \qquad (7)$$

Let us underline the principal features of the suggested approach.
1. We introduce a novel variable $G$ which can be used to derive additional information about a system in the stationary nonequilibrium state. We suppose that $G$ is a measurable quantity at macroscopic level, thus values like entropy which are related to the order parameter (principal macroscopic variable) can be defined. At the mesoscopic level the variable $G$ is introduced as a variable with operational characteristics of a random process slave with respect to the process describing the order parameter.
2. We suppose that thermodynamic forces γ related to the novel variable can be defined. One can introduce the "equations of state" $b(<E>,<G>)$, $g(<E>,<G>)$. Thus we introduce the mapping (at least approximate) of the external restrictions on the point in the plane $b, g$
3. We suppose that a "refined" structure factor $w(E,G)$ can be introduced which satisfies the condition $\int w(E,G=y)dy=w(E)$ (ordinary structure factor). This function (like $w(E)$) is the internal (inherent) property of a system. At the mesoscopic level we can ascribe to this function some inherent to the system (at given restrictions $(b_0,g_0)$) random process. The structure factor has the meaning of the joint probability density for the values $E,G$ understood as the stationary distribution of this process. Provided the "reper" random process for the point $(b_0,g_0)$, one can derive therefrom the shape of the structure function. If we model the dependence of the system potential of the order parameter by some potential well, the lifetime distribution within one busy period and probabilities $w(E,G)$ can be viewed as distributions of the transition times between the subset of the phase space (possibly of the fractal character) corresponding to the potential well, and the subset corresponding to the domain between the "zero" and the "hill" of the potential where from the system will roll down to the zero state. To determine the explicit form of $\tilde{A}$ (at $(b_0,g_0)$) the algorythm of the asymptotic phase coarsening of complex system is used (Section 3).
4. It is supposed that at least for certain classes of influences the resulting distribution has the form (5), (6), that is the change of the principal random process belongs to some class of the invariance leading to this distribution which explains how one can pass from the process in the reper point $(b_0,g_0)$ (for example, in equilibrium when $g=0$ and $b=1/k_BT$) to a system in an arbitrary nonequilibrium stationary. The thermodynamic forces should be chosen so that the distribution lead to new (measurable) values of $(<E>,<G>)$.

## 3. Distribution for nondisturbanced lifetime

If $g=0$ and $b=b_0=(k_BT_{eq})^{-1}$, where $k_B$ is the Boltzmann constant, $T_{eq}$ is the equilibrium temperature, then the expressions (5-6) yield the equilibrium Gibbs distribution. One can thus consider (5-6) as a generalization of the Gibbs statistics to cover the nonequilibrium situation. Such physical phenomena as the metastability, phase transitions, stationary nonequilibrium states are known to violate the equiprobability of the phase space points. The value $g$ can be regarded as a measure of the deviation from the equiprobability hypothesis.

Of coarse, the thermodynamic description itself is supposed to be already coarse-grained. To get the explicit form of the $G$ distribution we shall use the general results of the mathematical theory of phase coarsening of the complex systems [16], which imply the following distribution of the lifetime for coarsened random process:

$$P(G<y)=G_0^{-1}exp\{-y/G_0\} \qquad (8)$$

for one class of the stable states and the Erlang density

$$p(G<y)=\sum_{k=1}^{n} P_k G_{0k}^{-1}exp\{-y/G_{0k}\}; \qquad \sum_{k=1}^{N} P_k=1 \qquad (9)$$

in the case of several ($n$) classes of the ergodic stable states. The values $G_0$ and $G_{0i}$ are averaging of the residence times and the degeneracy probabilities over stationary ergodic distributions (in our case - Gibbs distributions). The physical reason for the realization of the distribution in the form (8-9) is the existence of the weak ergodicity in a system. Mixing the system states at big times will lead to the distributions (8-9). As we note in in Section 2, the structure factor $w(E,G)$ has a meaning of the joint probability density of values $E,G$. For the distributions (8),(9) the functions $w(E,G)$ from (4),(6) take on the form:

$$w(E,G)=w(E)G_0^{-1}exp\{-G/G_0\}; \quad w(E,G)=w(E)\sum_{k=1}^{n} P_k G_{0k}^{-1}exp\{-G/G_{0k}\} \qquad (10)$$

for the relations (8) and (9) respectively. Substituting into the partition function yield

$$Z(b,g)=Z(b)(1+gG_0)^{-1}; \quad Z(b,g)=Z(b)\sum_{k=1}^{n} P_k(1+gG_{0k})^{-1} \qquad (11)$$

for (8) and (9), where $Z(b)=\int w(E)exp\{-bE\}dE$ is the Gibbs partition function. For $n=2$, $Z(g)=P_1/(1+gG_{10})+P_2/(1+gG_{20})$; $G_{k0}=1/(-s_k))$, $k=1,2$, where $s_{1,(2)}=-(L_{11}+L_{22})/2+(-)[(L_{11}+L_{22})^2/4-L_{11}L_{22}+L_{12}L_{21}]^{1/2}$; $P_1=(G_{20}-G_{10}G_{20}b)/(G_{20}-G_{10})$; $P_2=(G_{10}-G_{10}G_{20}b)/(G_{10}-G_{20})$; $b=m_1(L_{22}+L_{21})+m_2(L_{11}+L_{12})$; $m_1+m_2=1$, $L_{kj}$ are the transition intensities between the classes of states; $m_k$ is the probability of the initial state belonging to the $k$-th ergodicity class.

We have from (6)-(7) and (10) when $\Gamma_g = -\partial \ln Z(b,g)/\partial g$   $\Gamma_0(V) = \Gamma_g(V)|_{g=0}$;

$$\Gamma_g = \Gamma_0/(1+g\Gamma_0), \qquad g = 1/\Gamma_g - 1/\Gamma_0, \qquad (12)$$

that is $g$ is the difference between the inverse lifetimes of the open system $1/\Gamma_g$ and the system without external influences $1/\Gamma_0$ which can degenerate only because of its internal fluctuations. The value $g$ is thus responsible for describing the interaction with the environment and its existence is the consequence of the open character of a system. When defining γ one should take into account all factors which contribute to the interaction between the system and the environment. If one denotes in (20) $x = g\Gamma_0$, then $x = x_1 + x_2 + ... + x_n$, where the value $x_i$ is determined by the flux labelled by the index $i$. From expressions (3) and (5) - (6) it is visible, that the value $g$ is connected to the entropy production $s$. From comparison with the Extended Irreversible Thermodynamics [17] it is possible to show that $x = g\Gamma_0 \approx q_a t_{0a}/aR$, where $q_a$ are characteristic currents in the system, $a$ is density of values $a$ which is transferred by a currents $q_a$, $R$ is the size of the system, $t_{0a}$ is time of degeneration of the system, time for which the system will pass with homeokinetic plateau in a degeneration state. Time $t_{0a}$ is expressed through times of a relaxation of currents $t_q$ (for example, for neutron system in a nuclear reactor with a current of neutrons $F$, $t_F = 1/v_n S_{tr}$, where $v_n$ is average speed of neutrons, $S_{tr}$ is transport section) and Onzager's factors $L_q$. If in the system there are sources of value $a$, $x_a = y_a t_{0a}$, $y_a = (q_a - R s_a)/Ra$, where $s_a$ is density of a source of value $a$.

At $n=1$ from correlations such as (11) it is possible to obtain the description of such nonequilibrium phenomena, as heat conductivity, mass transfer, the chemical reactions [4, 5], close to the description by means of the Extended Irreversible Thermodynamics [17].

Let's note, that the value similar $g$ it is defined in works [18-20] for fraktal object. It is equal to zero for the closed system, and for open system it is equal to $S l_i - l_{KS}$, where $l_i$ are Lyapunov's parameters, and $l_i$ is Kolmogorov-Sinai entropy.

## 4. Superstatistics from distribution of the kind (5).

In the distribution (5) containing lifetime, as thermodynamic parameter, probability for $E$ and $\Gamma$ is equal

$$P(E,\Gamma) = \frac{e^{-bE-g\Gamma} w(E,\Gamma)}{Z(b,g)}; \quad Z(b,g) = \int e^{-bE-g\Gamma} d\tau = \iint dE\, d\Gamma\, w(E,\Gamma) e^{-bE-g\Gamma}. \qquad (13)$$

Having integrated (13) on $\Gamma$, we obtain distribution of a kind

$$P(E) = \int P(E,\Gamma) d\Gamma = \frac{e^{-bE}}{Z(b,g)} \int_0^\infty e^{-g\Gamma} w(E,\Gamma) d\Gamma. \qquad (14)$$

According to the third assumption from Section 2, the structural factor $w(E,\Gamma)$ is meaningful to joint probability for $E$ and $G$, treated as stationary distribution of this process. We shall write down

$$w(E,\Gamma) = w(E)w_1(E,\Gamma) = w(E)\sum_{k=1}^{n} P_k f_k(\Gamma, E). \qquad (15)$$

In last equality (15) it is supposed, that there exists $n$ classes of ergodic states in a system; $P_k$ is the probability of that the system will be in $k$-th a class of ergodic states, $f_k(\Gamma, E)$ is density of distribution of lifetime $G$ in this class of ergodic states (generally $f_k$ depends from $E$). As physical example of such situation (characteristic for metals, glasses) one can mention the potential of many complex systems of a kind, for example, *Fig. 1*. Such situation is considered in [21]. Minima of potential correspond to the metastable phases, disproportionate structures, etc. Essential object of research of statistical physics recently became complex nonergodic systems: the spin and structural glasses, disorder geteropolimery, the granular media, transport currents, etc. [22]. The basic feature of such systems will be, that their phase space is divided into isolated areas, each of which corresponds to a metastable thermodynamic state, and the number of these areas exponential exceeds full number (quasi)particles [23]. The quasithermodynamical theory of structural transformations of alloy *Pd-Ta-H*, based on this model, is constructed in [24]. Expression (15) to the description of such systems is applicable.

We shall assume an obvious kind of distribution $f_k$, having chosen it as gamma-distribution

$$f_k(x) = \frac{1}{\Gamma(a_k)} \frac{1}{b_k^{a_k}} x^{a_k-1} e^{-x/b_k}, \quad x>0, \quad f_k(x)=0;\ x<0;\ \int_0^\infty e^{-g_k x} f_k(x)dx = (1+g_k b_k)^{-a_k}. \qquad (16)$$

($G(a)$ is gamma-function). We assume, that in $k$-th metastable area $g=g_k$. We shall choose $a_k=g/l_k$; $b_k=G_{0k}l_k/g$, $l_k$ is intensity of energy flow in the system (subsystem), equal in dynamical equilibrium of an output intensity [25]. Then $b_k a_k = G_{0k}$, $(1+g_k b_k)^{-a_k} = (1+l_k \Gamma_{0k})^{-g_k/l_k}$; $G_{0k}$ is equal to average lifetime in $k$-th a class of metastable states without disturbance, that is, in equilibrium. For average lifetime of the system in dynamical equilibrium state in work [25] by means of stochastic models of storage it is obtained

$$\Gamma_{ok} = \frac{1}{l_k}(Q_k - 1); \quad 1 + l_k \Gamma_{0k} = Q_k, \qquad (17)$$

where $Q_k = exp\{-b_k P_k V_k\}$ is the statistical sum of the grand canonical ensemble of the system in $k$-th a metastable state, $b_k=(1/k_B T)_k$ is return temperature in $k$-th a metastable state, $P_k$, $V_k$, $T_k$ are according to pressure, volume, temperature in $k$-th a metastable state, $V$ is the full system volume. The value $x=gG_0=q_a t_{0a}/aR$ is equal $g(Q-1)/l$, and $a=g/l= q_a t_{0a}/aR(Q-1)$. For systems of great volume $V$, $R \sim V^{1/3}$, $a<<1$. Values $a$ can be big for systems of small volume or with small lifetime, with the big currents $q_a$ and great values $t_{0a}$, at high temperature (and small $\mathbf{b}$ and $Q$), etc. From (16) follows, that value $\mathbf{g}_k$ from (5)-(7), (10)-(14), (16) which is thermodynamical conjugated to a thermodynamical variable $\mathbf{G}_k$ in $k$-th a metastable state, it is proportional to value $\mathbf{l}_k$, i.e.

$$\mathbf{g}_k = \mathbf{a}_k \mathbf{l}_k, \qquad (18)$$

where $\mathbf{a}_k$ is parameter of distribution (16). Then from (16) - (17) it is seen that

$$\int_0^\infty \exp\{-\mathbf{g}\Gamma\} f_k(\Gamma, E) d\Gamma = (1 + \mathbf{l}_k \Gamma_{0k})^{-\mathbf{a}_k} = Q_k^{-\mathbf{a}_k} = \exp\{-\mathbf{a}_k (P/k_B T)_k V_k\}; \quad \mathbf{a}_k = \mathbf{g}_k / \mathbf{l}_k. \qquad (19)$$

For exponential distribution $f_k$ in (16) $\mathbf{a}_k = 1$. From (13), (17) at $\mathbf{g}=1$ we obtain, that $<\mathbf{G}>=\mathbf{G}_0/Q=(Q-1)\mathbf{l}Q$. For the big systems, when $Q-1<<1$, $<\mathbf{G}>\sim 1/\mathbf{g}$ as in ordinary Boltzmann-Gibbs statistics, $<E>\sim 1/\mathbf{b}=k_B T$. At $\mathbf{g}^1 \mathbf{l}$, $<\mathbf{G}>=1/\mathbf{g}[1+(\mathbf{l}/\mathbf{g})/(Q-1)]\sim 1/\mathbf{g}$. Substitution (15), (19) in (14) gives

$$P(E) = \exp\{-\mathbf{b}E\} \mathbf{w}(E) Z^{-1} \sum_{k=1}^n P_k \exp\{-\mathbf{a}_k \ln(1 + \mathbf{l}_k \mathbf{G}_{0k})\} =$$

$$\exp\{-\mathbf{b}E\} \mathbf{w}(E) Z^{-1} \sum_{k=1}^n P_k \exp\{-\mathbf{b}_k(P_k v_k/u)\mathbf{a}_k E\}, \qquad (20)$$

where $u = E/V$ is specific energy, $v_k = V_k/V$. Having denoted $r_1 = P`v`\mathbf{a}`/u$, $\mathbf{a}_k \circledR \mathbf{a}`$, $P_k \circledR P`$, $v_k \circledR v`$, $\mathbf{b}_l = (1/k_B T`)$, and replaced summation by integration and entering density of probability $f(\mathbf{b}_l r_1) = \P P_k / \P (\mathbf{b}_l r_1)$, we obtain from (20):

$$P(E) = \exp\{-\mathbf{b}E\} \mathbf{w}(E) Z^{-1} \int_0^\infty f(\mathbf{b}_l r_1) \exp\{-\mathbf{b}_l r_1 E\} d(\mathbf{b}_l r_1); \quad \int_0^\infty f(y) dy = 1. \qquad (21)$$

For a case of one class ergodic states:

$P(E) = \exp\{-\mathbf{b}E\} \mathbf{w}(E) Z^{-1} \exp\{-\mathbf{b}_0(P_0/u)\mathbf{a}_0 E\} = \exp\{-\mathbf{b}_0 E\} \mathbf{w}(E) Z^{-1}$; $\mathbf{b}_0 = \mathbf{b} + \mathbf{b}_0(P/u)\mathbf{a}$; $\mathbf{b}_0 = \mathbf{b}/(1-\mathbf{a}P/u)$,

where $\mathbf{b}$ is the Lagrange parameter; value $\mathbf{b}_0 = 1/k_B T_0$ corresponds to average return temperature of full system; $k_B T_0$ characterizes the conveniently averaged energy; $r_0 = (P_0/u)\mathbf{a}_0 = (P/u)\mathbf{a}$ for full system; $\ln Q = \mathbf{b}_0 PV$; $\mathbf{b} = \mathbf{b}_0(1-\mathbf{a}P/u) = \mathbf{b}_0(1-r_0)$. Substituting this value $\mathbf{b}$ in (21) and replacing $\mathbf{b}_l$ by $\mathbf{b}_{0l} = 1/k_B T_k = 1/k_B T_l(r,t)$, the fluktuating value of temperature, we get:

$$P(E) = \exp\{-\mathbf{b}_0(1-r_0)E\} \mathbf{w}(E) Z^{-1} \int_0^\infty f(\mathbf{b}_{0l} r_1) \exp\{-\mathbf{b}_{0l} r_1 E\} d(\mathbf{b}_{0l} r_1) = \qquad (22)$$

$= f_B f_A / Z = P_B P_A (Z_B Z_A / Z); \quad Z = \breve{\phi} f_B f_A dE; \quad Z_A = \breve{\phi} f_A dE; \quad Z_B = \breve{\phi} f_B dE; \quad P_B = f_B / Z_B; \quad P_A = f_A / Z_A;$

$$f_B = exp\{-\pmb{b}_0(1-r_0)E\}\pmb{w}(E); \quad f_A = \int_0^\infty f(\pmb{b}_{01}r_1)exp\{-\pmb{b}_{01}r_1E\}d(\pmb{b}_{01}r_1),$$

where $f_A$ and $P_A$ corresponds to the type-A superstatistics from [26]. For $f(\pmb{b}_{01}r_1)=\pmb{d}(\pmb{b}_{01}r_1-\pmb{b}_0r_0)$ we obtain from (22) the ordinary Boltzmann factor. Equality $\pmb{b}_{01}r_1=\pmb{b}_0r_0$ it is possible at $\pmb{b}_{01}\gg\pmb{b}_0$, $P_k \gg P$, $l_k \gg l$. Then at $\pmb{g}_k \sim 1/\pmb{G}_{0k}$, $\pmb{g} \sim 1/\pmb{G}_0$ equality $\pmb{b}_{01}r_1=\pmb{b}_0r_0$ it is reduced to the correlation $\pmb{G}_{0k}/V_k=\pmb{G}_0/V$. As in [1], we can write down $P(E)=exp\{-\pmb{b}_0(1-r_0)E\}\pmb{w}(E)Z^{-1}exp\{-\pmb{b}_0r_0E\}[1+\pmb{s}^2E^2/2+O(\pmb{s}^3E^3)]=exp\{-\pmb{b}_0E\}\pmb{w}(E)Z^{-1}[1+(q`-1)(\pmb{b}_0r_0)^2E^2+g(q`)(\pmb{b}_0r_0)^3E^3+...]$, where $(q`-1)(\pmb{b}_0r_0)^2=\pmb{s}^2$, $q`=<(\pmb{b}_0r_0)^2>/<(\pmb{b}_0r_0)>^2$; $(q`-1)^{1/2}=\pmb{s}/(\pmb{b}_0r_0)$; $\pmb{s}^2=<(\pmb{b}_{01}r_1)^2>-(\pmb{b}_0r_0)^2$; value $\pmb{b}$ from [1] replace by $\pmb{b}_{01}r_1$, the fluktuating intensive parameter is equal $\beta_{01}r_1$ instead of $\pmb{b}$ as in [1]; $\pmb{b}_0r_0=<\pmb{b}_{01}r_1>=\int\pmb{b}_{01}r_1 f(\pmb{b}_{01}r_1)d(\pmb{b}_{01}r_1)$, function $g(q`)$ depends on the superstatistics chosen. For $f=f_G$ from (16), (23), $q`=q=q_{Ts}$ [2,3]. For others $f$: $q`=q_{BC}$ [1]. For any distribution $f(\pmb{b}_{01}r_1)$ with average $\pmb{b}_0r_0=<\pmb{b}_{01}r_1>$ and variance $\pmb{s}^2$ we can write $P(E) \sim exp\{-\pmb{b}_0(1-r_0)E\}<exp\{-\pmb{b}_{01}r_1E\}>=exp\{-\pmb{b}_0E\}<exp\{-(\pmb{b}_{01}r_1-\pmb{b}_0r_0)E\}>=exp\{-\pmb{b}_0E\}[1+\pmb{s}^2E^2/2+\sum_{r=2}^\infty (-1)^r<(\pmb{b}_{01}r_1-\pmb{b}_0r_0)^r>E^r]$ similarly [1].

If one performs the replacement of variables $\pmb{b}`=\pmb{b}_0(1-r_0)+r_1\pmb{b}_{01}$ and assumes that the situation $\pmb{b}_0=0$ is possible, when the bottom limit of integration after replacement of a variable is 0, instead of (22) we shall obtain that $P(E)=\pmb{w}(E)Z^{-1}\int_0^\infty f_1(\pmb{b}`)exp\{-\pmb{b}`E\}d\pmb{b}`$, $f_1(\pmb{b}`)=f(\pmb{b}_{01}r_1)=f[\pmb{b}`-\pmb{b}_0(1-r_0)]$, that coincides with superstatistics [1]. But the corerlation (22) describes a more general situations and superstatistics forms here a special case (22). The value $r_0=\pmb{a}_0(P_0/u)$. For ideal gas $P/u \approx 0,687$. Then $1<\pmb{a}_0<2$. Generally, for example, for liquids values $u<0$ and $P/u<0$ are possible. This relation and for other, more complex systems is known.

It is possible to mention some examples of physical realization of distribution (22). So, for a case of superdiffusion in [27] at the absence of external force stationary distribution is in Tsallis form, and for a case of presence of constant external force and multiplicate noise in exponential form of distribution is manifested. As in (22), in one phenomenon for one system two kinds of distribution are combined.

Besides the stated scheme there are also other opportunities of realizing the distributions such as (22) and generalizations of the theory of superstatistics. Except for distributions (15) - (16) for density of probability of lifetime in the structural factor $\pmb{w}(E,\pmb{G})$ (15) on can consider other distributions as well. Since the phase space of complex systems gets intricate fractal structure [28] it is natural to build a distribution on a fractal substrate. As distribution $f$ in (22) it is possible to choose a fractal distributions as in [21] or their generalizations on a case of multifractals, and also other distributions used in [1]. Other various combinations of different distributions in the different purposes are possible as well. The obtained distribution of a kind (22) gives the wide opportunities of the description of various physical situations.

For distribution function f in (22) as gamma-distribution such as (16) (but with variables $\pmb{a}_k=c$, $b_k=b$; $f(\pmb{b}_1r_1)=\pmb{G}^{-1}(c)b^{-c}(\pmb{b}_1r_1)^{c-1}exp\{-\pmb{b}_1r_1/b\}$) expression (22) reduces to the product of Gibbs distribution and Tsallis distribution of a kind

$$P(E)=exp\{-\pmb{b}_0(1-r_0)E\}\pmb{w}(E)Z^{-1}(1+bE)^{-c}=$$

$$=exp\{-b_0(1-r_0)E\}w(E)Z^{-1}[1+(q-1)b_0r_0E]^{-1/(q-1)}; \quad c=1/(q-1); \quad bc=b_0r_0. \qquad (23)$$

It is obvious that distribution (22) represents the product of Boltzmann factor with a multiplier superstatistik from which Tsallis distribution as a special case can be obtained (having chosen a $f$ in (22) the gamma-distribution of a kind (16)). Comparison of distribution (23) with Boltzmann-Gibbs distribution and with Tsallis distribution shows its intermediate behaviour: tails of distribution fall down not so quickly, as in Gibbs distribution, but also not so slowly, as in Tsallis distribution (*Fig. 2*). If for Gibbs distributions $l(u)=exp\{-b_0u^2/2\}/Z$ ($b_0=0,05$) tails on borders of the chosen area are equal $1,7 \cdot 10^{-10}$, and for Tsallis distribution $f(u)=[1+(q-1)b_0u^2/2]^{1/(1-q)}/Z$ ($q=1,46$) are equal about $3,7 \cdot 10^{-4}$, for distribution (23) $k(u)=exp\{-b_0(1-r_0)u^2/2\}Z^{-1}[1+(q-1)b_0r_0u^2/2]^{-1/(q-1)}$ ($r_0=0,8$) they are equal $7 \cdot 10^{-6}$ (and equal $1,004 \cdot 10^{-9}$ for $p(u)$ and $r_0=0,3006$).

The behaviour of distribution (22), (23) essentially depends on values $r_0$ ($a_0$). At $r_0 \to 0$ ($a_0 \to 0$, $a_0 > 0$) the Gibbs distribution is obtained, and at $r_0 \to 1$, $r_0 < 1$ one gets the Tsallis distribution. At $a \to 0$ and $g \to 0$ (18). As it is marked in the beginning of section 3, it corresponds to transition to Gibbs distributions. Thus and $l \to 0$, but also $a=gl \to 0$ also it is performed correlation (17). The value $g$ describes entropy fluxes and entropy production in the system; the value $l$ describes intensity of an input in system (and an output from it).

Distribution of a kind (23) was used by us for the description self-organized criticality (*SOC*) [29]. In work [30] formation of a stationary single avalanche and fluktuated formation of an avalanche in view of additive noise for a component of speed and an inclination of sand are considered. Thus distribution of the order parameter acquires a power-like form with an integer parameter. But generally this parameter should be fractional, therefore in [30] generalization of system of Lorentz is performed allowing to account for the behaviour of ensemble of avalanches. Stochastic systems for not additive ensemble of avalanches are used also. We have performed calculation on distribution (23) with effective energy

$$b_0E \to U(s)=ln[(I_S+I_Z s^t)/(1+s^t)^2]+\int_0^s [u-u^{t2}/(1+u^t)][(1+u^t)^2/(I_S+I_Z u^t)]du$$

from [30] for a case fluktuated formations of avalanches with an integer exponent of distribution of parameter of the order $t=2$ also have compared behaviour of stationary distribution to the results received in [30] (*Fig. 22* in [30]) for a case $t=1,5$ in distribution $exp\{-U(s)\}/Z$. The results are shown in *Fig.3*. Calculation was carried out for the same values of parameters noise intensivity of energy $I_S$ and complexity $I_Z$ which are used in [30]. Concurrence of the results, similar behaviour of distributions is evidenced. Parameters $q$, $r_0$ only weakly influence behaviour of distribution. We shall note, that using the pure Tsallis distribution leads to another behaviour of stationary distribution (in particular, does not show indicated in [30] distinctions between distributions with $I_S=1$; $I_Z=5$ and $I_S=0,5$; $I_Z=30$).

Comparison with the results received in works [31, 32] for a spectrum of the cosmic ray was carried out as well. By means of distribution of a kind $p(E)=CE^2(1+(q-1)\tilde{b}E)^{-1/(q-1)}$ ($C$ is a constant representing the total flux rate; the Tsallis distribution is multiplied with $E^2$, taking into account the available phase space volume) at $q=1,215$, $\tilde{b}^{-1}=107$ $MeV$, $C=5 \cdot 10^{-13}$ in [31] the measured flux rate of cosmic ray particles with a given energy is well fitted. If we use for these purposes the distribution (23) multiplied with $CE^2$ there is an additional parameter $r_0$ (*Fig.4*). At values $r_0$ close to $1$ we get coincidence to results of [31]. At $r_0=1-10^{-14}$ the coincidence to results [31] is exact. At $r_0=1-10^{-13}$ sharper bend is obtained at ankle energy $\sim 10^{19}$ eV; the plot for flux rate comes nearer to curve for parameters $q=11/9$, $C=10^{-14}$ from [31] as well. Decreasing $r_0$ we get the dependence, characteristic for Boltzmann-Gibbs distribution, resulted on *Fig*.1 in [32]. At values $r_0 \approx 1$ it is possible to explain conformity of experimental data in small volume of system both high energy and temperatures. Comparison of calculations was carried out also by means of expression (23) with estimations of distributions for the motion of point defects in thermal convection patterns in an inclined fluid layer [26], [33] which also has shown conformity with the results received in [26], [33].

## 5. Conclusion

From the distribution (5) containing lifetime of statistical system (time before degeneration of system) are obtained the superstatistics, entered in [1] in many respects is formal (though in [1] the dynamic bases is lead). In the present work obvious expression for $\beta=\beta_0(1-r_0)+r_1\beta_{01}$ is obtained, the physical sense of distributions $f_1(\beta)$ and $f(\alpha_k,\beta_k,P_k/u_k)$ is elucidated (besides distribution of return temperature), - as density of distribution of probability of destruction in the certain time interval, on $k$-th stages. It is possible for this reason gamma-distributions $f(x)=\lambda^a x^{a-1} exp\{-\lambda x\}/\Gamma(a)$ with a parameter $a=r/2$ (is more true, $c^2$ distributions with $r$ degrees of freedom which corresponds to gamma-distribution at $\lambda=1/2$-$c$; in [1] value $r$ is interpreted as number of degrees of freedom giving the contribution in fluktuating value $\beta$) precisely describe Tsallis statistics, - they correspond to distribution of lifetime with $r$ stages.

Other variant of the obtained expressions forms at $lnQ=\beta PV$ instead of $\beta_0 PV$ as it was supposed after expression (21). Then $\beta_0=\beta(1+r_0)$, $\beta=\beta_0/(1+r_0)$, and correlations (22), (23) become the form $P(E)=exp\{-\beta_0 E/(1+r_0)\}w(E)Z^{-1}\int_0^\infty f(\beta_{01}r_1/(1+r_0))exp\{-\beta_{01}r_1E/(1+r_0)\}$ $d(\beta_{01}r_1/(1+r_0))$; $P(E)=exp\{-\beta_0 E/(1+r_0)\}w(E)Z^{-1}[1+(q-1)\beta_0 r_0 E/(1+r_0)]^{-1/(q-1)}$. At $r_0 \to 0$, $\alpha \to 0$, these distributions pass in Gibbs distribution, as in (22), (23). But in Tsallis distribution they pass at $r_0 \to \infty$, $\alpha \to \infty$, $\gamma \to \infty$ as against (22), (23) where it occurs at $r_0=\alpha P/u=1$, $\alpha=u/P=\gamma\lambda$. Possible relation of values $\beta_q=\beta'\sum_{j=1}^W p_j^q =\beta'[1-(q-1)S_q]$ [34] with $\beta_0=\beta(1-r_0)$ or $\beta(1+r_0)$ depends on a choice of $\beta$ as well. Thus it is possible to find a relation between $\alpha$ and $q$. In [34] one more interesting question on a form of the differential equations for $y(x)=P(E)$ is considered. For distribution (23) this equation looks like $dP(E)/dE=-\beta_0(1-r_0)P(E)-\beta_0 r_0 P(E)/[1+(q-1)\beta_0 r_0 E]$. It is possible to lead analogies to the equations which have been written down in [34].

The Gibbs distribution does not describe the dissipative processes that develop in the system. Superstatistics describe systems by constantly putting energy into the system which is dissipated. The value $\alpha=\gamma\lambda$ is connected with dissipative processes in the system (through $\gamma$). She defines a correlation between Gibbs and superstatistics multipliers in distribution (22).

It is possible not to pass to integrated relations, as in (22), having limited to summation (so, as discrete analogue of gamma-distribution negative binomial distribution serves). The discrete description in many cases, for example, for bistability potential, appears more precisely continuous. The found conformity between superstatics and the nonequilibrium distribution containing lifetime, should appear useful to both cases. For example, many results established for superstatistics and nonextensive statistical mechanics, are transferred to the description of complex systems by means of the distribution containing lifetime. Since thermodynamics with lifetime [4,5] is more general, than the theory of superstatistiks also it has more opportunities. Interesting is establishing the relation to a method of the nonequilibrium statistical operator of Zubarev [11] generalizing Gibbs distributions which as it was marked, it is possible to compare to the distributions containing lifetime, and nonextensive statistical mechanics [2-3], in which entropy is represented by means of the measures which are distinct from Boltzmann and Gibbs. Probably, they stand close to the concept of lifetime, in the latent kind present in both theories. The relation $\boldsymbol{g}=\boldsymbol{al}$ (18) is of interest also. The value $\boldsymbol{g}$ thermodynamical conjugated to random lifetime is expressed through entropy production and currents [12], i.e. through communication of system with an environment.

It is shown, that the behaviour of the obtained distribution (22)-(23) interpolates between behaviour of Gibbs and Tsallis distributions. Application of this distribution to the phenomenon of the self-organized criticality (and other examples) shows his efficiency. The obtained distribution contains the new parameter related to a thermodynamic state of the system, and also with distribution of a lifetime of a metastable states and interaction of this states with an environment. Changing this parameter it is possible to pass to both Gibbs and Tsallis distributions.

## Figure captions

Fig.1. Schematic dependence of thermodynamic potential on energy in a case potential when the phase space of system is divided into isolated areas, each of which answers a metastable thermodynamic state.

Fig.2. Comparison of distributions (23) $k(u)=exp\{-\boldsymbol{b}_0(1-r_0)u^2/2\}Z^{-1}(1+(q-1)\boldsymbol{b}_0 r_0 u^2/2)^{-1/(q-1)}$ ($r_0=0,8$) and $p(u)$ ($r_0=0,3006$) with Boltzmann-Gibbs distribution $l(u)=exp\{-\boldsymbol{b}_0 u^2/2\}/Z$ ($\boldsymbol{b}_0=0,05$) and with Tsallis distribution $f(u)=[1+(q-1)\boldsymbol{b}_0 u^2/2]^{1/(1-q)}/Z$ ($q=1,46$).

Fig.3. Function of distribution (23) in the description self-organized criticality (SOC) at $\boldsymbol{t}=2$, $q-1=4/11$, $r_0=0,1$; $P(s)$ $(R(s), f(s))=[1+0,1(q-1)U(s)]^{-1/c}exp\{-0,9U(s)\}$; $U(s)=ln[(I_S+I_Z s^t)/(1+s^t)^2]+\int_0^s [u-u^{t2}/(1+u^t)][(1+u^t)^2/(I_S+I_Z u^t)]du$; $P(s): I_S=0; I_Z=50$; $R(s): I_S=0,5; I_Z=30$; $f(s): I_S=1; I_Z=5$.

Fig.4. Energy spectrum of primary cosmic rays (in units of $m^{-3}s^{-1}sr^{-1}GeV^{-1}$) as listed in [31]. The line $p(E)$ is the prediction by $p(E)=CE^2(1+(q-1)\tilde{\boldsymbol{b}}E)^{-1/(q-1)}$ with $q=1,215$, $\tilde{\boldsymbol{b}}^{-1}=k_B\tilde{T}=107$ MeV, $C=5\cdot 10^{-13}$ in the above units [31]. The function $f(E)=CE^2 exp\{-\tilde{\boldsymbol{b}}(1-r_0)E\}(1+(q-1)\tilde{\boldsymbol{b}}r_0 E)^{-1/(q-1)}$ obtained on (23) with $r_0=1-10^{-14}$. Functions $g(E), k(E), r(E), s(E), e(E)$ too are obtained on expression (23) with $r_0=1-10^{-13}$, $r_0=1-10^{-12}$, $r_0=1-10^{-11}$, $r_0=1-10^{-10}$, $r_0=1-10^{-9}$ accordingly.

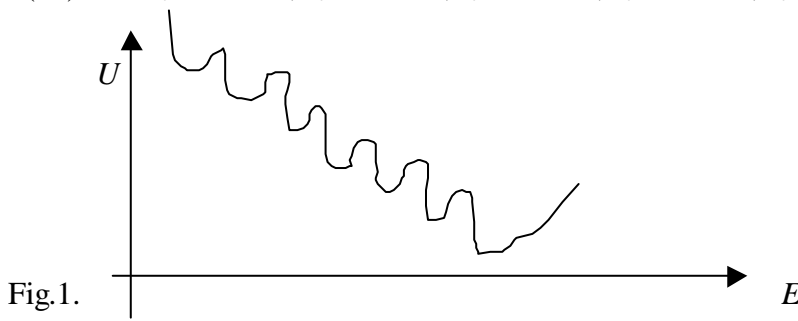

Fig.1.

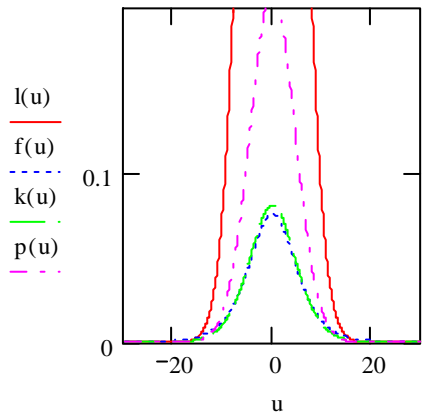

Fig.2.

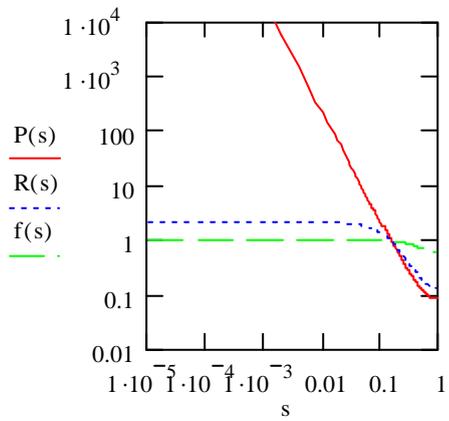

Fig.3.

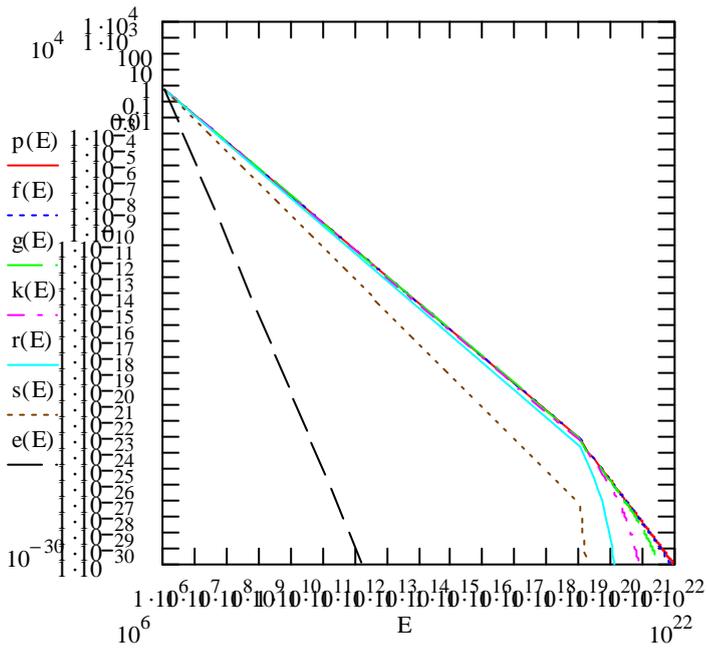

Fig.4.